\documentclass[fleqn,usenatbib]{mnras}

\usepackage{newtxtext,newtxmath}

\usepackage[T1]{fontenc}
\usepackage{ae,aecompl}

\usepackage{graphicx}	
\usepackage{amsmath}	
\usepackage{amssymb}	

\title[Decoupling the rotation of stars and gas - II]{Decoupling the rotation of stars and gas - II: the link between black hole activity \& MaNGA kinematics in TNG}

\author[C. Duckworth et al.]
{Christopher Duckworth,$^{1,2}$\thanks{E-mail: cd201@st-andrews.ac.uk}
Tjitske K. Starkenburg,$^{2}$ 
Shy Genel,$^{2}$ 
Timothy Davis,$^{3}$ \newauthor
M\'elanie Habouzit,$^{5,6}$
Katarina Kraljic,$^{4}$
Rita Tojeiro$^{1}$ 
\\
$^{1}$School of Physics and Astronomy, University of St Andrews, North Haugh, St Andrews, KY16 9SS, UK\\
$^{2}$Flatiron Institute, 162 Fifth Avenue, New York, NY 10010, USA\\
$^{3}$School of Physics \& Astronomy, Cardiff University, Queens Buildings, The Parade, Cardiff, CF24 3AA, UK\\
$^{4}$Institute for Astronomy, University of Edinburgh, Royal Observatory, Blackford Hill, Edinburgh EH9 3HJ, UK\\
$^{5}$Max-Planck-Institut fur Astronomie, Konigstuhl 17, D-69117, Heidelberg, Germany\\
$^{6}$Zentrum f\"ur Astronomie der Universit\"at Heidelberg, Institut f\"ur Theoretische Astrophysik, D-69120 Heidelberg, Germany\\
}

\date{Accepted XXX. Received YYY; in original form ZZZ}

\pubyear{2019}

\begin{document}
\label{firstpage}
\pagerange{\pageref{firstpage}--\pageref{lastpage}}
\maketitle

\begin{abstract}
We study the relationship between supermassive black hole (BH) feedback, BH luminosity and the kinematics of stars and gas for galaxies in IllustrisTNG. We use a sample of galaxies with mock MaNGA observations to identify kinematic misalignment at $z=0$ (difference in rotation of stars and gas), for which we follow the evolutionary history of BH activity and gas properties over the last 8 Gyrs. Misaligned low mass galaxies typically have boosted BH luminosity, BH growth and have had more energy injected into the gas over the last 8 Gyr in comparison to aligned galaxies. These properties likely lead to outflows and gas loss, in agreement with active low mass galaxies in observations. We show that splitting on BH luminosity at $z=0$ produces statistically consistent distributions of kinematic misalignment at $z=0$, however, splitting on the maximum BH luminosity over the last 8 Gyrs does not. While instantaneous correlation at $z=0$ is difficult due to misalignment persisting on longer timescales, the relationship between BH activity and misalignment is clear. High mass quenched galaxies with misalignment typically have similar BH luminosities, show no overall gas loss, and have typically lower gas phase metallicity over the last 8 Gyrs in comparison to those aligned; suggesting external origin.
\end{abstract}

\begin{keywords}
galaxies: kinematics and dynamics -- galaxies: active -- galaxies: evolution
\end{keywords}



\section{Introduction}
In the framework of a $\Lambda$ cold dark matter Universe, galaxies form from the cooling and condensation of the initial gas cloud within dark matter haloes \citep{white1978, mo1998}. In the basic picture, they inherit the angular momentum content of the surrounding halo \citep[][]{fall1980}, which is obtained in the early growth phase by tidal torques from the large-scale structure \citep[e.g.][]{peebles1969, Doroshkevich1970}. As stars form from the rotating gas, they exhibit its dynamical characteristics often leading to coherent rotation between dark matter, gas and stars in both magnitude and direction. However, after turnaround there is good reason to believe that the rotation of dark matter, gas and stars may decouple from each other as galaxies evolve up-to $z=0$. 
 
Recent cosmological scale hydro-dynamical simulations have provided a clear insight into the relationship between the angular momentum of baryons and dark matter through cosmic time. A necessary component of realistic simulations is efficient feedback from both supermassive black holes (BH) and stars, required to, amongst other things, reproduce late type disks and solve the problem of catastrophic angular momentum loss \citep[e.g.][]{zavala2008, scannapieco2009}. Active galactic nuclei (AGN) and supernova explosions can also lead to dramatic redistribution of gas which regulate the angular momentum content of galaxies \citep[e.g.][]{genel2015, DeFelippis2017}. 

`Quasar' (radiative) mode feedback releases huge amounts of energy through radiation from the accretion disk leading to high luminosity AGN and dramatic gas outflows \citep[e.g.][]{cattaneo2009, rubin2014, cheung2016}. Alternatively `radio' (kinetic) mode is termed for lower luminosity AGN that host lower black hole accretion rates. In this instance energy is deposited into the surrounding gas via jets which drive outflows, heat the gas and suppress star formation \citep[][]{binney1995, ciotti2001, heckman2014}.

The relationship between AGN and kinematics has been the focus of several recent studies using Integral Field Spectroscopy (IFS) data. In particular, a potential new class of galaxy termed `red geyser' has been identified which host AGN and exhibit high velocity outflows in the spatial distribution of ionized gas \citep[][]{cheung2016, roy2018}. These outflows are often linked to a distinctive offset in rotation direction between the stars and gas. Detection of ongoing outflows is, however, rare ($\sim5-10$\% of the quiescent population). \citet{penny2018} demonstrate the importance of AGN feedback in low mass quiescent galaxies ($\mathrm{M_{stel} < 5 \times 10^{9}M_{\odot}}$). While the majority demonstrate no ionized gas present, quiescent galaxies (i.e. with reduced or null SFR) with an AGN show a clear decoupling in the rotation of stars and gas. However, the relationship of gas kinematics to BH feedback is not clear for all galaxies \citep[see also:][]{koudmani2019}. In particular, \citet{ilha2019} find that the typical decoupling between stars and gas for AGN defined galaxies are consistent with an inactive control sample. 

Termed kinematic misalignment, the decoupled rotation of stars and gas can also be a natural result of external processes \cite[e.g.][]{davis2011, barrera2015, vdvoort2015, jin2016, bryant2019, duckworth2019_halo}. Regardless of internal or external origin, kinematic misalignment in observations and simulations is linked with both a lower gas mass fraction and angular momentum \citep[][]{duckworth2019}. \citet{starkenburg+19} highlight the importance of feedback leading to gas loss, enabling re-accretion as a mechanism for future misalignment in low mass galaxies. The question arises if misalignment is caused in the first instance by mergers (and hence making BH accretion easier) or if they are a result of AGN feedback leading to gas loss. 
The timescales of luminous AGN are typically much shorter than kinematic misalignment, making correlation at $z=0$ alone difficult. 

In this letter, we study the temporal relationship between BH feedback, BH luminosity and kinematic misalignment in the cosmological scale hydrodynamical simulation of IllustrisTNG100 (hereafter referred to as TNG100). We use a sample of galaxies with mock MaNGA \citep[Mapping Galaxies at Apache Point;][]{bundy2015, blanton2017} observations at $z=0$ to emulate what we may expect to see in IFS observations. In Section \ref{sec:methods} we briefly describe the simulation and how we construct our sample. In Section \ref{sec:results} we present our results before concluding in Section \ref{sec:conclusion}.

\section{Methods} \label{sec:methods}
We use the fiducial run of TNG100 which follows the evolution of a periodic cube with side lengths of 110.7 Mpc (75 h$^{-1}$ Mpc) with mass resolution of baryonic (dark matter) elements of 1.4 x 10$^6 \mathrm{M_{\odot}}$ (7.5 x 10$^6 \mathrm{M_{\odot}}$). The IllustrisTNG project \citep{marinacci18,naiman18,nelson18,pillepich18b,springel18} is a suite of magneto-hydrodynamic cosmological scale simulations incorporating an updated comprehensive model for galaxy formation physics \citep[][]{weinberger17,pillepich18a} and making use of the moving-mesh code \texttt{AREPO} \citep{springel10,pakmor11,pakmor13}. We make use of public data, as described in \citet{nelson2019}. Of particular relevance is the prescription of \citet{weinberger17} for BH feedback, which is modelled by two modes (quasar mode: high accretion, kinetic mode: low accretion). The transition threshold in terms of the Eddington ratio is $\mathrm{f_{Edd}= \min ( 2x10^{-3}(M_{BH}/10^8 M_{\odot})^2 , 0.1)}$ so that BHs typically transition from quasar to kinetic mode around $\mathrm{M_{BH} = 10^{8}M_{\odot}}$. 


To emulate what we may expect to see in IFS observations we construct a sample of TNG100 galaxies with mock MaNGA observations \citep[a complete description is given in][]{duckworth2019}. Here we briefly describe the sample construction. 
For each public MaNGA galaxy, we find an unique object in TNG100 with the closest match in stellar mass, $g-r$ colour and size. MaNGA is designed have a near flat distribution in stellar mass ($\mathrm{10^{8.5} M_{\odot} < M_{stel} < 10^{11.5} M_{\odot}}$). We create mock observations for each match by convolving the raw motions of all star particles/gas cells with noise and PSF typical of MaNGA. Each galaxy is `observed' (taking the line of sight to the be the z-axis of the box) up-to the typical MaNGA footprint (1.5-2.5 effective light radii in a distinct hexagonal shape) to create mock velocity fields, from which we define the degree of misalignment between the rotation of stars and gas by fitting a position angle (PA). This is done using the \texttt{fit\_kinematic\_pa} routine \citep[see Appendix C of][]{krajnovic2006} so that $\Delta\mathrm{PA = |PA_{stellar} - PA_{gas}|}$. We take objects with $\Delta$PA < 30$^{\circ}$ to be aligned, $\Delta$PA $ \geq 30^{\circ}$ to be misaligned and $\Delta$PA $\geq 150^{\circ}$ to be counter-rotating. We only select galaxies that have distinct PAs (i.e. clear, coherent rotation) in both their stellar and gas velocity fields, leaving a total of $\sim$2500 galaxies used in this study. We compute BH bolometric luminosities as:
\begin{equation}
\mathrm{L_{bol, AGN}} = \frac{\varepsilon_r}{1 - \varepsilon_r} \mathrm{\dot{M}_{BH} c^2}
\end{equation}
where $\varepsilon_r=0.1$ is the radiative efficiency \citep[see discussion in][]{habouzit2019}, c the light speed, and $\mathrm{M_{BH}}$ the accretion rate onto the BH. Gas properties are defined within two effective radii ($\mathrm{R_{e}}$, radius containing half of the stellar mass within the galaxy), unless stated otherwise.

\vspace{-1em}
\section{Results} \label{sec:results}
\begin{figure}
	\includegraphics[width=0.95\linewidth]{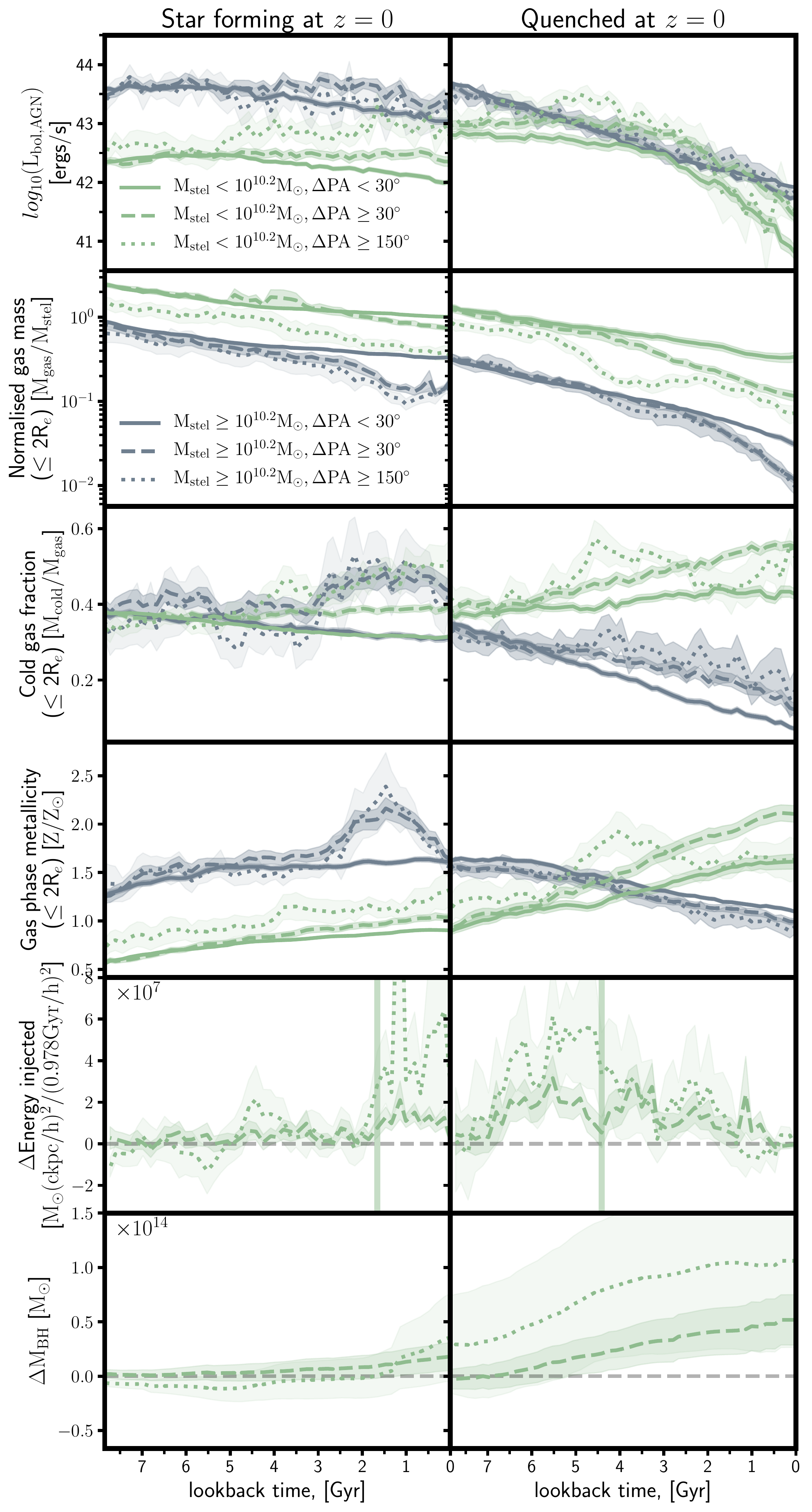}
    \caption{Time evolution of (rows top to bottom) black hole luminosity ($\mathrm{log_{10}(L_{bol, AGN})}$), normalised gas mass ($\mathrm{M_{gas}/M_{stel}}$), cold and star forming gas fraction ($\mathrm{M_{cold}/M_{stel}}$), gas phase metallicity, rate of energy injection from feedback and black hole mass for star forming (left) and quenched galaxies (right) identified at $z=0$. For the top 5 rows, we show both low mass (green; $\mathrm{M_{stel} < 10^{10.2}M_{\odot}}$) and high mass (grey; $\mathrm{M_{stel} > 10^{10.2}M_{\odot}}$) galaxies. Both are subdivided by misalignment ($\Delta$PA $< 30^{\circ}$: solid, $\geq 30^{\circ}$: dashed, and  $\geq 150^{\circ}$: dotted). Each line shows the mean for the population with the shaded region corresponding to the standard error. The bottom two rows (black hole energy injection and black hole mass) are residuals for the low mass galaxies only for $\Delta$PA $\geq 30^{\circ}$ (dashed) and $\Delta$PA $\geq 150^{\circ}$ (dotted) defined relative to $\Delta$PA$ < 30^{\circ}$. The vertical line in energy injection panels represents the time at which 50\% of the energy over the last 8 Gyrs has been injected.}
    \label{fig:overall_pop}
\end{figure}

Each panel of Figure \ref{fig:overall_pop} shows the time evolution average of a property for all galaxies split by stellar mass at $\mathrm{M_{stel} = 10^{10.2}M_{\odot}}$. This corresponds to the typical transition from quasar to kinetic feedback \citep[i.e. $\mathrm{M_{BH} \approx 10^{8}M_{\odot}}$, see Fig 1 in][]{li2019}. We note that splitting in this way ensures that we isolate quasar feedback for the low mass sample, however, our high mass sample has been subject to both quasar and kinetic feedback over the last 8 Gyrs. Splitting on BH mass or enforcing stricter stellar mass cuts does not change any of our findings. We divide our sample by $\Delta$PA at $z=0$. For each sub-population, we find that the stellar mass distributions are consistent at $z=0$ for aligned and misaligned galaxies. We also split on specific star formation rate using the distance from the star-forming main sequence as defined in \citet{pillepich2019}. We select star forming ($\Delta \mathrm{log_{10}(SFR) > −0.5}$) and quenched galaxies ($\Delta \mathrm{log_{10}(SFR) \leq -1.0}$) at $z=0$.

For low mass galaxies, BH luminosity (top row; $\mathrm{log_{10}(L_{bol, AGN})}$) is distinctly higher for the misaligned (and counter-rotating) at $z=0$ over the last 8 Gyr compared to those aligned. In contrast, we note that the correlation between misalignment and BH luminosity for the high mass bin is less clear. Given that low mass galaxies only exhibit quasar mode feedback, this suggests that this feedback mode is more strongly correlated with misaligned gas rotation. 

Gas loss (second row) is a key feature for all misaligned galaxies within 2$\mathrm{R_{e}}$, potentially indicative of gas outflows from the BH feedback. Prior to the gas loss there is an increase in the fraction of gas (third row) that is defined as cold ($\mathrm{T < 10^{4.5}K}$) or star forming (SFR > 0) which could suggest that inflow of new material triggers the high accretion state and future feedback. The higher gas phase metallicity (fourth row) for misaligned low mass galaxies could indicate that the potential inflow is re-cycled material or that fresh accretion is prevented by radiative outflows.

Conversely, for the quenched high mass galaxies the metallicity is typically lower for those misaligned at $z=0$. In addition we find that despite gas loss within 2$\mathrm{R_{e}}$, the misaligned show typically higher cold gas masses (within 2$\mathrm{R_{e}}$) and no overall gas loss (within the overall subhalo) relative to the aligned (supplementary material; rows one and two). This may indicate that accretion of pristine gas and gas rich minor mergers is more important for decoupling their rotation. This could be a natural result of quenched galaxies hosting smaller gas reservoirs, meaning that only a small amount of accretion (relatively to late-types) leads to misalignment. An alternate explanation could be that enriched gas is preferentially lost due to feedback or environment. The angular momentum content of the gas is disrupted to a similar degree as for the low mass galaxies (supplementary material; third row).

For the misaligned low mass galaxies, the trends of BH luminosity and gas loss are more prominent and have deviated from the aligned galaxies at earlier times for those quenched at $z=0$. To understand this, in the bottom two panels we show the time evolution of the energy injection into the surrounding gas cells and black hole mass (see supplementary material for high mass equivalent of these panels). Here, we plot the average residual difference of misaligned (dashed) and counter-rotating (dotted) galaxies with respect to the aligned galaxies (grey dashed line at the origin). 

\begin{figure*}
	\includegraphics[width=0.9\linewidth]{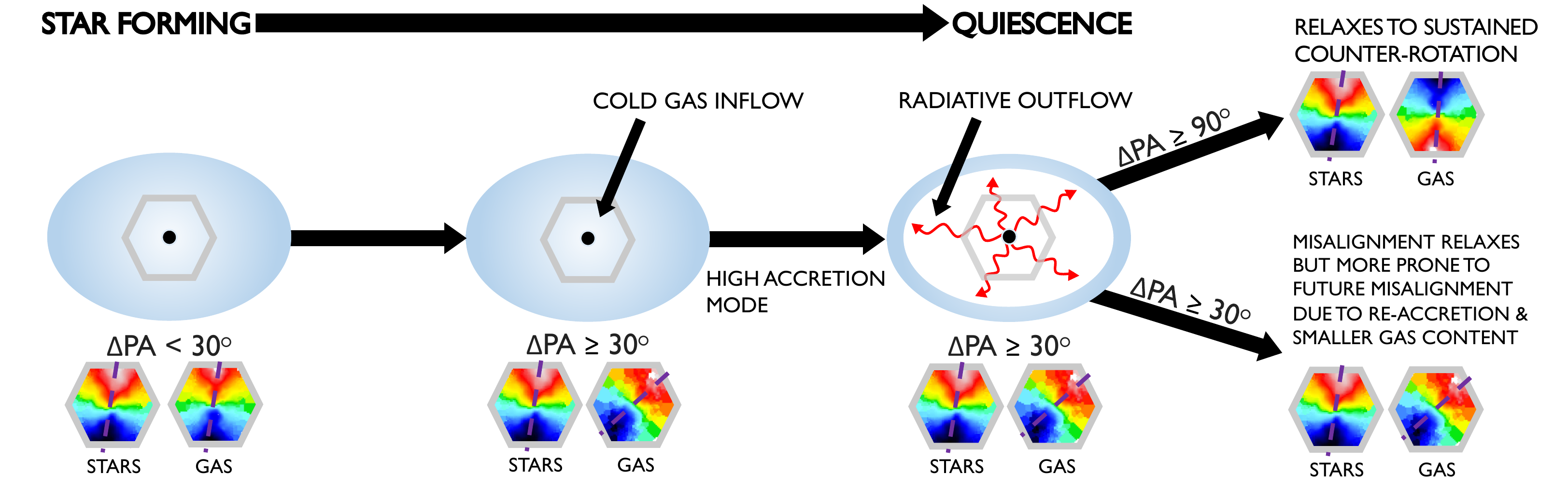}
    \caption{Diagram of the relationship between BH accretion, radiative feedback, and kinematic misalignment in low mass galaxies. The diagram (left to right) shows the time evolution of radiative feedback which injects thermal energy into the cold gas reservoir of galaxy after a high accretion state triggered by a possible inflow of cold gas. Below this are representations of stellar and gas velocity fields showing what we may expect to see in an IFS observation if it was made at this time. $\Delta$PA refers to the difference between the global position angles (purple lines) for the stellar and gas velocity fields. High accretion onto the BH is directly related to boosted BH luminosity in our model. Over time radiative outflows act to suppress star formation leading to quiescence. Given that first misalignment ($\Delta$PA $\geq 30^{\circ}$) most commonly occurs around the high accretion period, we see that selecting different sSFR low mass galaxies at $z=0$ would result in different timescales since the peak AGN luminosity. Star forming galaxies with misalignment at $z=0$ will have had more recent peak energy injection since the feedback hasn't had time to fully suppress star formation yet (fifth row in Figure \ref{fig:overall_pop}).}
    \label{fig:diagram}
\end{figure*}

We find that the rate of energy injection is typically elevated for counter-rotating galaxies, however a clear boost can be seen for all those that are misaligned. The lookback time of peak energy injection is earlier for quenched galaxies relative to the star forming (the solid vertical line represents the time at which 50\% of the energy over the last 8 Gyrs has been injected). Misaligned star forming galaxies show far more recent feedback and BH luminosity, possibly indicative that gas is in the process of being decoupled and blown out, but has not acted to fully suppress star formation yet. Conversely, the quenched galaxies exhibit earlier peak energy injection which in turn suppresses star formation leading to the quenched classification at $z=0$. For this reason selecting misaligned star forming and quenched galaxies at $z=0$ will naturally lead to different time correlations in BH activity. This can be visualized by the diagram in Figure \ref{fig:diagram}. It is important to note that despite the increased fraction of cold or star forming gas for the misaligned galaxies, the overall gas mass is lower than the aligned.


We also show the residual time evolution of black hole mass with respect to the aligned galaxies. We find that BH growth correlates with the time scales of energy injection, indicating the close relationship between feedback and accretion for those misaligned. The causality between BH growth and feedback is however not clear. While BH growth leads to increased feedback, the question remains whether the angular momentum is disrupted prior to feedback (potentially due to mergers) or if feedback leading to gas loss (making it easier for future (re-)accretion to disrupt angular momentum) which then leads to increased BH growth.

\begin{figure}
    \centering
	\includegraphics[width=\linewidth]{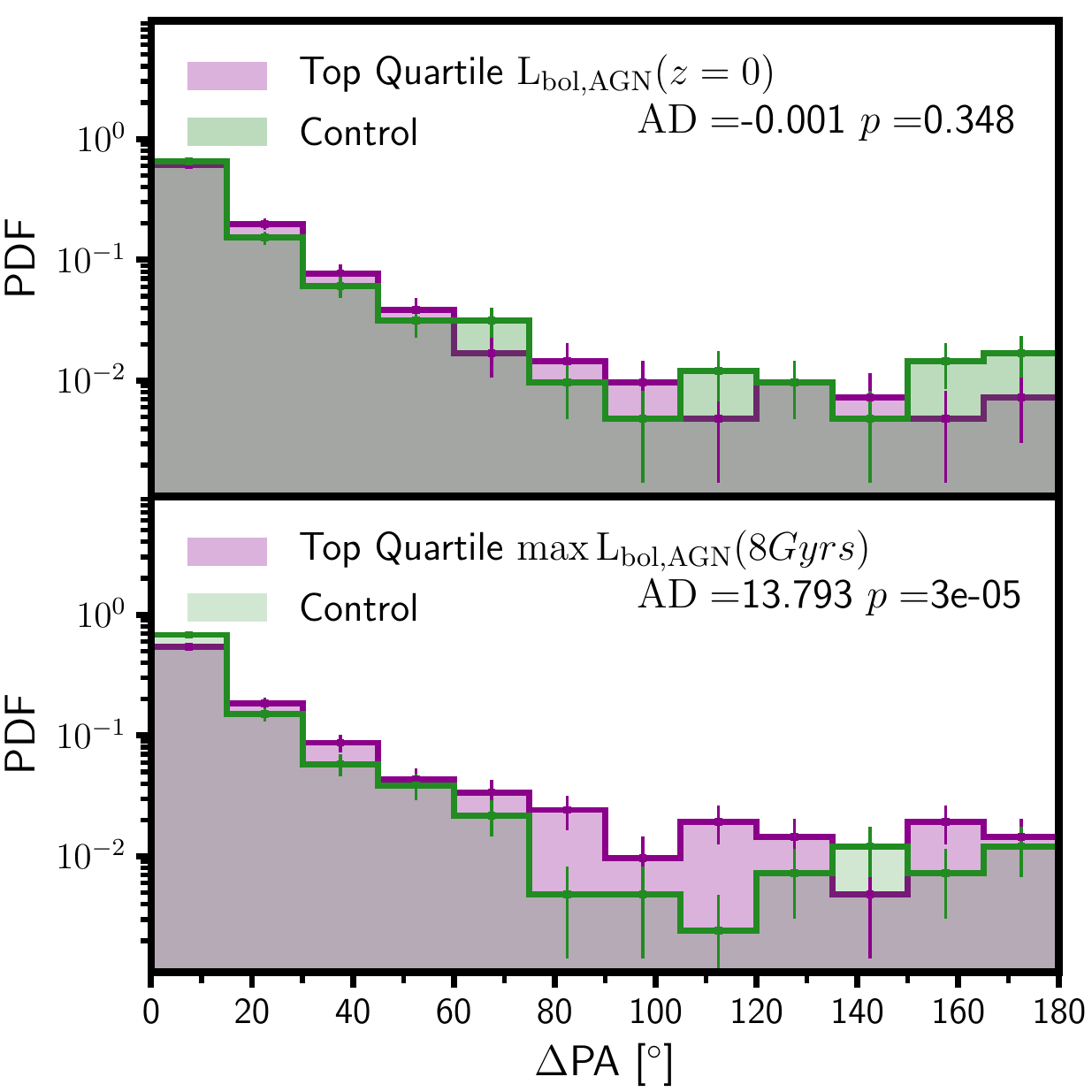}
    \caption{Probability density function of kinematic misalignment as defined by $\Delta$PA at $z=0$. In both panels the brightest 20\% in $\mathrm{L_{bol,AGN}}$ (purple) compared with a mass matched control (green) is shown. The top panel shows the brightest in $\mathrm{L_{bol,AGN}}$ at $z=0$ only, whereas the bottom panel shows those with the brightest peak $\mathrm{L_{bol, AGN}}$ over the last 8 Gyrs. In each panel the Anderson-Darling statistic with a corresponding p-value is shown. We find no statistical difference between the active galaxies and the control for those selected at $z=0$ only, whereas we find that those which have been the most luminous over the last 8 Gyrs are distinctly more misaligned.}
    \label{fig:PAdist}
\end{figure}

To understand how kinematic misalignment may correlate with BH luminosity at $z=0$ alone, in the top panel of Figure \ref{fig:PAdist}, we show the distribution of $\Delta$PA for the top 20\% BH luminosity in our low mass sample ($\mathrm{M_{stel} < 10^{10.2}M_{\odot}}$) in comparison with a control sample (all defined at $z=0$). The control is made by taking the closest unique match in stellar mass for each high BH luminosity galaxy from the remainder of our sample. We find the two distributions are statistically indistinguishable (Anderson-Darling statistic; -0.001 with a p-value of 0.348). In the bottom panel we show the same but instead we select the top 20\% in peak BH luminosity (for each galaxy) over the last 8 Gyrs. In this instance the AGN bright galaxies are distinctly more misaligned than the mass matched control (Anderson-Darling statistic; 13.793 with a p-value of $3 \times 10^{-5}$). 

This demonstrates that despite the inherent relationship between BH luminosity, feedback and gas kinematics, considering the overall distribution of $\Delta$PA split on instantaneous luminosity at $z=0$ does not necessitate that correlation is found. This matches that of observations in MaNGA \citep[Figure 6 in][]{ilha2019}, who also find no correlation between active galaxies and a control sample. We emphasise that while feedback is clearly correlated with misalignment, the decoupled rotation of gas can remain for several Gyr after initially becoming misaligned, meaning that a single time-step is unlikely to characterise the relationship for an ensemble of galaxies. 

We note that IllustrisTNG typically under-produces bright AGN ($\mathrm{L_{X-ray}(2-10 keV) > 10^{44}ergs^{-1}}$) for $z \leq 1$ in contrast with observational constraints \citep[see][]{habouzit2019}. Given this and the uncertainty in estimating BH luminosity from simulations (treatment of radiatively efficient and inefficient AGN \& obscuration), we chose to select by percentile rather than cutting on absolute luminosity. Regardless, selecting only bright AGN in this way or choosing a higher percentile does not change our conclusions.

\section{Summary} \label{sec:conclusion}

In this paper, we study the relationship between BH luminosity, BH feedback and kinematic misalignment between stars and gas for galaxies in TNG100. We use mock observations of an IFS survey (MaNGA) built from galaxies in TNG100, to identify kinematic misalignment ($\Delta$PA; difference in PAs of stars and gas) at $z=0$. We split our mock IFS sample on stellar mass to separate the impact of `quasar' and `kinetic' feedback modes. We follow the time evolution of BH luminosity and energy injection from BH feedback in leading up to misalignment (or counter-rotation) at $z=0$. We also compare the $z=0$ distributions of $\Delta$PA of the most luminous BHs in our sample against a control. Our conclusions are as follows.
\begin{enumerate}
    \item Low mass galaxies with misalignment (and counter-rotation) at $z=0$ typically have had boosted BH luminosity, BH growth and significantly more energy injected into the gas over the last 8 Gyr in comparison to aligned galaxies. Gas is potentially blown out due to the AGN feedback, losing angular momentum towards $z=0$. Prior to the feedback there is an increase in the fraction of cold phase gas within 2$\mathrm{R_{e}}$ (seen also for high mass galaxies), along with an increased metallicity. This is seen for all populations split on $z=0$ sSFR.
    
    \item The epoch of peak energy injection from the quasar mode feedback is different as a function of $z=0$ sSFR for misaligned low mass galaxies. This can be explained by the relationship between energy injection from feedback and galaxy quenching (see Figure \ref{fig:diagram}). Misaligned quenched galaxies have typically experienced peak energy injection from the BH at earlier times which has since acted to suppress star formation, whereas misaligned star forming galaxies exhibit more recent energy injection. 

    \item Quenched high mass galaxies with misalignment (and counter-rotation) at $z=0$ typically have similar BH luminosity over the last 8 Gyr with respect to aligned galaxies and gas loss is only seen within 2$\mathrm{R_{e}}$ (but not for the total galaxy). Gas phase metallicity is also lower with respect to aligned galaxies. This suggests that the origin of misalignment in massive quenched galaxies is more likely due to accretion of pristine gas or loss of enriched gas. 
    
    \item We find that the distributions of kinematic misalignment are statistically indistinguishable between the top 20\% in BH luminosity of low mass galaxies in our sample and a mass matched control at $z=0$. This matches observations \citep[see Figure 6 in][]{ilha2019}. Misalignment may initially occur at a similar time to the initial high accretion state (and hence peak BH luminosity), however misalignment can persist/correlate on much longer timescales. To test this we split by the top 20\% in maximum BH luminosity (for each galaxy) over the last 8 Gyrs in comparison with a control. We find that the most luminous AGN over the last 8 Gyrs are significantly more misaligned at $z=0$. This result suggests that while one may not expect correlation with misalignment when considering BH activity at $z=0$ alone, the relationship between BH luminosity and misalignment in low mass galaxies is clear. 
\end{enumerate}


\section*{Acknowledgements}
CD acknowledges support from the Science and Technology Funding Council (STFC) via an PhD studentship (grant number ST/N504427/1). The Flatiron Institute is supported by the Simons Foundation. We thank the IllustrisTNG team; this work was conducted using the public data release, however benefited from use of the private data for other research. 

\vspace{-1em}

\label{lastpage}
\bibliographystyle{mnras}
\bibliography{biblio} 


\end{document}